\documentclass[twocolumn,showkeys,preprintnumbers,amsmath,amssymb]{revtex4}

\usepackage{graphicx}
\usepackage{natbib}
\usepackage[T2A]{fontenc}
\usepackage{dcolumn}
\usepackage{bm}
\usepackage{hyperref}
\usepackage{float}
\restylefloat{table}
\newcommand{\RNumb}[1]{\uppercase\expandafter{\romannumeral #1\relax}}

\begin{document}

\title{Precision measurement of gravitational frequency shift of radio signals using Rao-Cramer estimates}

\author{Aleksei V. Belonenko}
\email{av.belonenko@physics.msu.ru}
\author{Andrei V. Gusev}%
\author{Valentin N. Rudenko}%
\email{valentin.rudenko@gmail.com}

\affiliation{%
	Sternberg Astronomical Institute, Moscow State University, Moscow, 119991 Russia 
}%

\date{\today}

\begin{abstract}
	\mbox{Abstract} -- A method has been developed for precision measurement of the gravitational frequency shift of communication radio signals between the spacecraft and the ground tracking station based on the maximum likelihood principle, using the Rao-Cramer limit estimates for the kinematic parameters associated with orbital motion. Numerical illustrations of the efficiency of the method are presented using the example of data obtained in experiments with the Spectr-R satellite as part of the VLBI system in the ``Radioastron'' mission. A compensatory ``on-line'' technique for suppressing Doppler and atmospheric noise has been implemented due to the presence of two modes of communication in gravity sessions: unidirectional (1w) and looped (2w). Recipes for reducing the magnitude of systematic errors are discussed.
\end{abstract}
\maketitle	
	\section[Introduction ]{Introduction}
The problem of measuring the effect of gravitational redshift using the Spektr-R satellite as part of the Radioastron (RA) mission was considered in \cite{linkk1}, \cite{linkk2}. In these articles, it was assumed that the advantage of the possibility of multiple measurements arising from the cyclic orbital motion of the spacecraft provides an increase in accuracy proportional to the square root of the number of measurements performed in convenient orbits. Thus, the presence of H-maser standards on board and on the ground tracking station (GTS) of the same quality as in the GP-A experiment \cite{linkk3}, can help to achieve an accuracy of an order of magnitude better than provided for in the GP-A mission; i.e. at the level of  $10^{-5}$  after 100 data accumulations in orbit.
Testing the effect of gravitational redshift has recently been the subject of research with the Galileo 5 and 6 navigation satellites \cite{linkk4},\cite{linkk5}, which had slightly elongated elliptical orbits with eccentricity $\sim 0.1$. The trajectory tracking of these devices was carried out by the European VLBA radio telescope system and partially by laser ranging stations. According to the authors of the works, as a result of data processing using estimated compensation algorithms, they managed to show the validity of the Einstein formula for the gravitational frequency shift between ground and onboard standards with an accuracy of  $(3.6)\times 10^{-5}$, i.e., to improve the GP-A result to three times.
The more specialized ACES misson \cite{linkk6},\cite{linkk7}, which is expected to launch on the ISS (The International Space Station) in the coming years, has the goal of achieving an accuracy of $2 \times 10^{-6}$.
The total duration of the operation of the Spektr-R satellite was about 7 years from the launch in 2011 to 2018. Special gravity sessions of approximately an hour duration were held from 2015 to 2018, when the onboard hydrogen frequency standard finished working.
In total, about 70 sessions have been accumulated, which were processed in the first approximation. ``Single-track data'' using synchronization of communication signals according to the onboard standard were processed in a large volume (more than 5,000 sessions, since all sessions were involved, not just specialized gravity ones) and the results are published in  \cite{linkk8}. 	
The correspondence of the measured value ``redshift'' to the general theory of relativity formula is confirmed at the level of  $10^{-2}$. At this level, it is not necessary to take into account and eliminate subtle effects such as the influence of atmospheric interference \cite{linkk9}. Compensation was mainly subject to the Doppler frequency shift. Processing of gravity sessions with switching of synchronization modes between ``one-way'' and ``two-way'' was performed using the so-called ``webinet data'', obtained using standard spectral frequency meters at the tracking station of the ``Puschino''. The accuracy of the frequency measurement of this equipment is limited by the relative error $\delta f/f = 10^{-13}$. The processing results give the boundary of the correspondence of the measured value ``redshift'' to Einstein's theory improved by an order of magnitude, i.e. at the level of $10^{-3}$ \cite{linkk10}. 	
The rejection of the use of ``webinet'' and the transition to optimal methods for estimating the frequency of communication signals, in principle, should allow us to compare the measurements even more accurately with the general theory of relativity formula \cite{linkk11}. It is expected that with the correct optimization of the procedure for processing records of gravitational sessions, the accumulated data will be enough to test the correspondence of theory and experiment with an accuracy of  $10^{-4} \div 10^{-5}$, i.e. at the level of the results obtained in measurements with Galileo satellites \cite{linkk4},\cite{linkk5}.
The purpose of this article is to develop such a procedure within the maximum likelihood criterion using the Rao-Kramer limit estimates under conditions of a narrow band of communication signals and large SNR.	
	
\section{Pre-processing of records: spectral frequency estimation algorithm}	
The physical and technical characteristics of the mission of the RA spacecraft were previously presented in several articles. A detailed description can be found in the \cite{linkk12}. Carrying a parabolic antenna, the Spectr-R satellite had a very eccentric elliptical orbit around the Earth, varying from cycle to cycle (due to the gravitational influence of the Moon and other factors) in a wide range of parameters: the perigee height from 1000 to 80,000 km, and the apogee height from 270,000 to 350,000 km. The orbital period ranges from 8 to 10 days. The shift of the gravitational frequency between the clocks on the Earth's surface and those infinitely distant has the order  $\Delta f/f \approx 6 \times 10^{-10}$. For onboard clocks, their shift changes along the orbit: the amplitude modulation reaches $0.6$  in orbits with low perigee $(1\div10) \times 10^3 $ km and less than 10 times in most medium orbits with perigees $\sim 50 \times 10^3 $ km.	
On board the SC and GTS (Ground Tracking Station), two identical hydrogen frequency standards were installed with a minimum deviation of the Allen variance of  $2 \times 10^{-15}$ for the averaging time of $\sim3600 \ s$ («Vremya CH», [16]). Two main carrier frequencies were used in the GTS communication line with the spacecraft: 8.4 GHz and 15 GHz. The first is the so-called ``raw tone'' for technical control and tuning, and the second is the transmission of scientific data about astrophysical objects detected by a space radio telescope. In principle, both frequency channels are suitable for measuring gravitational redshift. However, such measurements are more effective with a special communication session, the so-called ``combined communication mode'', which in most cases is incompatible with radio astronomy observations.
The quality of the measurement of the orbit parameters is characterized by the following values: the accuracy of the coordinates is 200 m. with radio control, but $\sim 2 $ cm. with laser ranging. The accuracy of the speed measurement is $\sim 2 \  \frac{mm}{s}$.	
A time-limited quasi-harmonic signal from the satellite is received on the GTS, converted to an intermediate frequency and then digitized.  
 The frequency measurement procedure is performed ``online'' on GTS by standard spectral meters, but a posteriori it is possible to obtain a more accurate estimate by recording the signal at the end of the observation time (an example of such a procedure can be found in \cite{linkk5},\cite{linkk11}). 
A typical empirical algorithm contains a number of transformations between the frequency and time domains. Usually, the signal spectrum is first constructed, and then the region of the maximum frequency component is filtered by the spectral window and shifted to the low-frequency side. Returning to the time domain reproduces the slow evolution of the phase over the observation interval; the phase drift is estimated by the LSM (least squares method), resulting in a regression curve. After subtracting it from the full phase, ``residuals'' are obtained, called the ``stopped phase''. The derivative of the stopped phase and its spectrum give an estimate of the frequency value and the ``line width''; that is errors in the frequency estimate. The most complete algorithm (and codes) of this type, which were used in our experiments, was developed at the JIVE Institute, which unites the European VLBI network of radio telescopes. In the following, we refer to the procedure for such data processing as ``carrier estimation'' by the JIVE algorithm ``swspectrometer'' and ``sctracker''.
Two main modes of operation were provided for the communication line of the Spectr-R satellite with GTS. The first is the ``H-maser'' or ``one-way'' the mode that is used to receive satellite signals at the carrier frequency with phase synchronization according to the onboard hydrogen standard. The second, ``coherent'' or ``two-way'' the mode is one in which the carrier is synchronized according to the GTS ground standard. The signal is sent to the satellite from the ground station, and then relayed back by the satellite without changing the synchronization (although changing the carrier value is possible). The presence of both communication modes provides a unique opportunity to filter the gravitational frequency shift from effects of a different nature. In particular, from the prevailing Doppler shift (which is four orders of magnitude greater). In a two-way communication signal, the Doppler shift is twice as large as in a one-way, and there is no gravitational frequency shift there completely \cite{linkk3}). 
Let's move on to the analytical description of the proposed filtering. To do this, we define the observed variables: the phase and its time derivative (frequency), which are formed from the input signal recorded on the GTS.	
Let’s consider a mixture of signal and noise at the input of the receiver
$$y(t) = S(t,\mathbf{a}) + n(t), \quad 0<t<T,$$
where
$$S(t,\mathbf{a}) = A\cos[\omega_0t+\varphi(t,\mathbf{a})]  = Ae^{j(\omega_0t+\varphi(t,\mathbf{a}))},$$
$$ \ \varphi(t,\mathbf{a}) = \sum_{k}a_kt^k \ -$$
$S(t,\mathbf{a})$ is a useful narrowband signal depending on the vector parameter $\mathbf{a} = ||a_0,\ ...\ , a_m ||^{\mbox{T}}$, and 
$$<n(t)n(t+\tau)> = N\delta(\tau) - \mbox{gaussian white noise}$$ 
Let the $\hat{a}_i = \gamma_i(y)$ be an unbiased estimate of an unknown parameter $a_i, \ i=\overline{0, M}.$ 
One can write down the expression for its variance:
\begin{equation}
\left\langle (a_i - \hat{a}_i)^2 \right\rangle \geq \sigma^2_{ii},
\end{equation}	
where $\sigma^2_{ii}$ - diagonal elements of the matrix $\vec{I}^{-1}, \ \vec{I} = [I_{ij}]$ - Fischer information matrix:
\begin{equation}\label{Fisher}
I = 
\begin{Vmatrix}
I_{11} & I_{12} & I_{13}\\
I_{21} & I_{22} & I_{23}\\
I_{31} & I_{32} & I_{33}\\
\end{Vmatrix} \rightarrow 
I^{-1} = 
\begin{Vmatrix}
\mathbf{\sigma_{11}^2} & \cdot & \cdot \\
\cdot & \mathbf{\sigma_{22}^2} &\cdot \\
\cdot & \cdot & \mathbf{\sigma_{33}^2}\\
\end{Vmatrix} 
\end{equation} 
$$I_{ij} = - \left\langle \frac{\partial^2\ln\Lambda(y|\mathbf{a})}{\partial a_i \partial a_j}\right\rangle = \frac{1}{N}\int_{0}^{T}\frac{\partial S(t,\mathbf{a})}{\partial a_i}\frac{\partial S(t,\mathbf{a})}{\partial a_j}dt \simeq $$
\begin{eqnarray}\label{0}
\simeq\frac{A^2}{2N}\int_{0}^{T} \frac{\partial \varphi(t,\mathbf{a})}{\partial a_i}\frac{\partial \varphi(t,\mathbf{a})}{\partial a_j}dt = \\ q\frac{T^{i+j}}{(i+j+1)}, \ i,j = 1,2,3,
\end{eqnarray} 	
where $\Lambda(y|\mathbf{a})$ - the conditional likelihood ratio, SNR, is denoted as $q$:
\begin{equation}
\ln\Lambda(y|\mathbf{a}) = \frac{1}{N}\int_{0}^{T}y(t)S(t,\mathbf{a})dt - \frac{1}{2}q,
\end{equation}
$\mbox{SNR} = q = \frac{A^2T}{2N}$. 
Maximum likelihood estimate $\mathbf{\hat{a}} = ||\hat{a}_0,\ ...\ , \hat{a}_m ||^{\mbox{T}}$ asymptotically effective: 
\begin{equation}
\lim_{q\rightarrow\infty}<(a_i - \hat{a}_i)^2> = \sigma^2_{ii}, \quad i = \overline{0, M},
\end{equation}
where $a_i$ are determined by a system of equations:

\begin{equation}\label{Ln}
\left[\frac{\partial\ln\Lambda(y|\mathbf{a})}{\partial a_i} \right]_{a_i=\hat{a}_i} = 0,\quad i = \overline{0, M}.
\end{equation}	
When receiving a phase-modulated signal, the maximum likelihood equations turn out to be nonlinear, which significantly complicates the technical implementation of such an algorithm. In practice, when measuring the instantaneous frequency 
\begin{equation}\label{+}
\omega(t)=\omega_0 + \Delta\omega(t), \quad \Delta\omega(t) = \dot{\varphi}(t, \mathbf{\hat{a}}),
\end{equation}	
some spectral estimation algorithms have become widely used.
In spectral estimation, the observation interval (0, T) is divided into short subintervals $(t_k, t_k + \Delta t), \ k = \overline{0, Q}$:
$$
y(t) \rightarrow \sum_{k}y_k(t),\quad y_k(t) = y(t_k \le t \le t_k+\Delta t).
$$
The spectrum is estimated for each subinterval by the FFT algorithm, followed by averaging over parallel observation channels, if this is possible. The sampling scale over the spectrum can be made sufficiently detailed with an artificial increase in the signal assignment interval, with the expansion area covered with zeros [14].
Hence, taking into account (\ref{Ln}), we have:
$$
\ln\Lambda(y|\mathbf{a}) = \sum_{k}\ln\Lambda(y_k|\mathbf{a}),
$$
\begin{equation}\label{Rud}
\ln\Lambda(y_k|\mathbf{a}) = \frac{A}{N}\mbox{Re} \int_{0}^{\Delta t}y_k(t)\exp\left\lbrace-j\Phi(t_k+t) \right\rbrace dt - \frac{\mbox{q}}{2},
\end{equation}
where
$$
\Phi(t) = \omega_0t + \varphi(t,\mathbf{a}),\ \mbox{q} =  \frac{A^2\Delta t}{2N}.
$$  
It can be seen from (\ref{Rud}) that the logarithm of the likelihood ratio on each k-time interval is proportional to the spectrum component $y_k(\omega)$ on this segment.

At short intervals, $(t_k, t_k+\Delta t)$ in a linear approximation, we represent:
\begin{equation}\label{^}
\Phi(t_k + \Delta t) \simeq \Phi(t_k) + \dot{\Phi}(t_k)\Delta t.
\end{equation}
In accordance with the theory \cite{linkk13}, a most plausible estimate of the instantaneous frequency $\omega(t)$ at the moment $t = t_k$, is the frequency at which the spectral component reaches its maximum at this interval:
\begin{equation}
\left| y_{k,\omega}(\omega_0 + \Delta\omega_k)\right| = max\left|Y_{k,\omega(\omega)}\right|,
\end{equation}
where $Y_{k,\omega}(\omega) \leftrightarrow Y_k(t)$ - spectrum of a random process $y_k(t)$.
For a large SNR $ \gg1$ the variance of the estimate $\Delta\hat{\omega}_k$ (\ref{+}) intends to the lower Rao-Cramer bound given by the diagonal elements of the Fisher inverse matrix \cite{linkk14}. In particular, the frequency variance looks like (see (\ref{0})):
\begin{equation}
\sigma^2_\omega \approx \frac{6}{q_k\Delta t^2}.
\end{equation}

When the influence of the receiver's own noise can be ignored, the error is determined by the frequency sampling step $\delta\omega$:
$$\delta\omega \gg \sigma_\omega$$ 
and then one should take
\begin{equation}
\left\langle (\omega_k - \hat{\omega}_k)^2 \right\rangle \approx \delta\omega^2 + \sigma^2_\omega \approx \delta\omega^2.
\end{equation}
Another source of systematic error in the spectral estimation algorithm is the frequency variations caused by the non-uniform motion of the spacecraft. Then a linear approximation of the phase evolution (\ref{^}) is not enough, it is necessary to take the following approximation:
\begin{equation}
\Phi(t_k + t) \approx \Phi(t_k) + \dot{\Phi}(t_k)t + \frac{1}{2}\ddot{\Phi}(t_k)t^2.
\end{equation}
	In this case, the useful signal $S_k(t,\vec{a})$ is considered linearly frequency-modulated and the Fisher matrix becomes more complicated.   
To eliminate the error caused by frequency modulation, a phase detection method can be applied using the algorithm of optimal frequency discriminator. In this algorithm the initial spectral estimation of the received signal acts as the first approximation in the iterative filtering procedure.
\section{Optimal discriminator for estimating kinematic parameters}	
\begin{figure}[h!]
	\centering
	\includegraphics[width=1\linewidth]{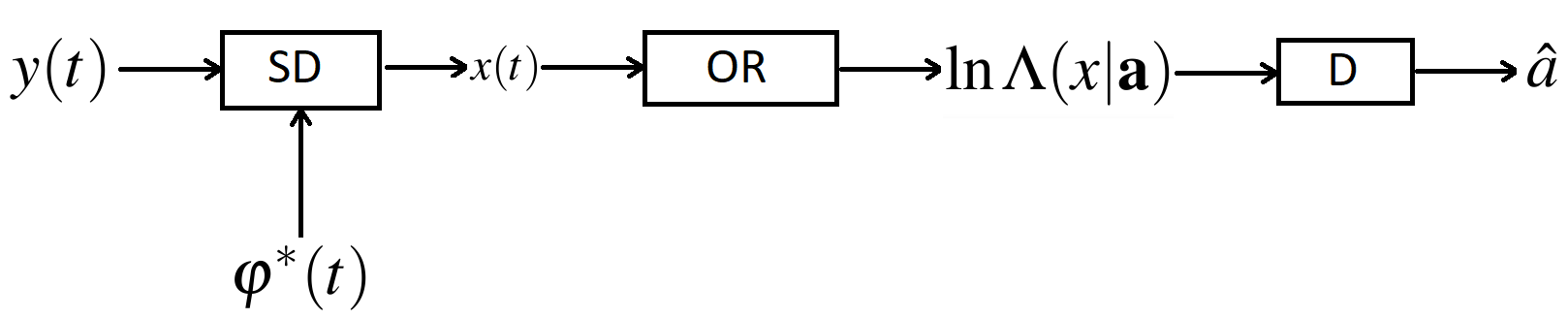}
	\caption{SD is a synchronous detector, OR is an optimal receiver with the output signal proportional to the $\sim\ln\Lambda(x|\mathbf{a})$, D - decision block with the  structure determined by a system of maximum likelihood equations.}
	\label{fig:1}
	\centering
\end{figure}
The most general algorithm for estimating the informative parameter $\varphi(t,\mathbf{a})$ is based on parallel processing in time of the implementation of the random process  $y(t)$ using a multichannel meter. An alternative approach, the advantage of which is in the technical implementation, consists in a sequential search, the main element of which is the discriminator of the monitored signal. Sequential search can be considered as a method of sequential approximations for solving a system of nonlinear maximum likelihood equations.

Let,
$$\varphi^*(t,\mathbf{a}) = \sum_{k}a^*_kt^k   $$
 a reference signal that is close enough to the estimated parameter $\varphi(t,\mathbf{a})$. Then: 

$$y(t) = A \cos\left[ \omega_0t + \varphi^*(t) +\Delta\varphi(t,\mathbf{a}) \right] + n(t), \ 0<t<T,$$
where
$$\Delta\varphi(t,\mathbf{a}) = \varphi(t,\mathbf{a}) - \varphi^*(t).$$
It is assumed that $(|\Delta\varphi(t,\mathbf{a})| \ll 1)$, as a first approximation, we have:
$$y_0(t) = -A\sin\Phi^*(t)\Delta\varphi(t,\mathbf{a}) + n(t), \quad 0<t<T.$$
Here
$$y_0(t) = y(t) - A\cos\Phi^*(t), \quad \Phi^*(t) = \omega_0t+\varphi^*(t),$$
$$\Delta\varphi(t,\mathbf{a}) = \varphi(t,\mathbf{a}) - \varphi^*(t) = \sum_{k}\Delta a_kt^k,$$
where
$$\Delta  a_k = (a_k - a_k^*), \quad $$ $a_k^*$ - the initial approximation coefficients obtained. For example,  with the JIVE algorithm (on the of stage ``swspectrometer'') or with the ``webinet'' frequency meter\footnote{webinet.asc.rssi.ru}.
The logarithm of  likelihood ratio for the random process $y_0(t)$ is defined by the following expression:
$$\ln\Lambda(y_0|\mathbf{a}) \approx \frac{(-1)A}{N}\int_{0}^{T}y_0(t)\sin\Phi^*(t)\Delta\varphi(t,\mathbf{a})dt -$$
$$ - \frac{A^2}{2N}\int_{0}^{T}\sin^2\Phi^*(t)\Delta\varphi^2(t,\mathbf{a})dt.$$
But in the further analysis, we will use a complex form of recording narrow-band process:

$$y(t) = \mbox{Re}\left[\tilde{y}\exp\left\lbrace j\omega_0 t\right\rbrace \right],$$
where 
$$\tilde{y}(t) = R(t)\exp \left\lbrace j\theta (t)\right\rbrace, $$
is the complex envelope with amplitude. 
Removing the terms at twice frequency $2\omega_0$, one comes to:

$$\ln\Lambda(y_0|\mathbf{a}) = $$
\begin{eqnarray}\label{sharp1}
\ln\Lambda(x|\mathbf{a}) =
\frac{A}{N}\int_{0}^{T}x(t)\Delta\varphi(t,\mathbf{a})dt -  \frac{A^2}{4N}\int_{0}^{T}\Delta\varphi^2(t,\mathbf{a})dt,
\end{eqnarray}
where
\begin{equation}\label{sharp2}
x(t) = R(t)\sin\left[ \theta(t) - \varphi^*(t)\right].
\end{equation}
Taking into account that
$$\Delta\varphi(t,\mathbf{a}) = \sum_{k}\Delta a_kt^k,$$
the system of maximum likelihood equations looks as
\begin{equation}\label{sharp3}
\frac{\partial\ln\Lambda(x|\mathbf{a})}{\partial\Delta a_i}\bigg|_{\Delta a_i = \Delta\hat{a}_i} = 0, \quad i = \overline{0,M},
\end{equation}
it can be represented in a matrix form:
$$\sum_{k}I_{ik}\Delta\hat{a}_k = F_i,$$
where $I_{ik}$ - elements of the Fischer information matrix, see (\ref{0}).
$$F_i = \frac{1}{2N}\int_{0}^{T}x(t)t^idt.$$
From here 
$$\Delta\hat{\mathbf{a}} = \mathbf{I}^{-1}\mathbf{F}, \qquad \mathbf{F} = ||F_0...F_M||^{T},$$
and, therefore,
$$<\Delta\hat{a}_k> = \Delta a_k = a_k - a^*_k,$$
$$<(\Delta\hat{a}_k - \Delta a_k)^2> = \sigma^2_{ii}.$$
In the first approximation in $\Delta\varphi(t,\mathbf{a}) \ll 1$
the estimates of the vector parameter $\mathbf{a}$ can be shifted $(\Delta a \neq 0);$ their variances are determined by the diagonal elements of the inverse Fisher matrix $\mathbf{I}^{-1}$. However, for large SNR, these variances tend to the minimum admissible Rao-Cramer boundary. 

In the next step, the discriminator operation can be repeated, substituting estimates of the first approximation as the vector of parameters. In fact, such an iterative process can be considered as a method of successive approximations when solving nonlinear equations of maximum likelihood. According to the scheme:
$$\varphi^*_{m+1}(t) = \sum_K\hat{a}_{k,m}t^k,$$
where $\hat{a}_{k,m}$ - estimates of unknown parameters $a_k$ at the m-step ($a^*_{k,m+1} = \hat{a}_{k,m}$).

At large signal-to-noise ratios, these estimates tend to be effective, i.e. unbiased, with variance determined by the lower Rao-Cramer boundary, already at the first iterations.

\section{Phase detector}
As can be seen from the expression (\ref{sharp1}),  $\ln \Lambda(x|a)$ is proportional to the correlation integral of the observed variable $x(t)$ which is the complex envelope of the input mixture $y(t)$ (in this case, the phase of the envelope $\theta(t)$ is reduced (compensated) for the phase of the initial approximation $\varphi^*(t)$, see (\ref{sharp2})). The difference between the polynomials of the current and seed phases is used as a reference signal (template). The most plausible estimates of the coefficients of the phase polynomial are obtained as solutions of the extremum equations (\ref{sharp3}) for corrections to the parameters $\mathbf{a}_i$. This procedure, often called a ``sequential discriminator'', proves the possibility of reaching extremely accurate estimates of the signal parameters corresponding to the Rao-Cramer boundary. However, if additional a priori information is available, it can be replaced by a simpler phase detection procedure. Such information, in particular, is the knowledge of a narrow received signal bandwidth and a large SNR.

As indicated in sections \RNumb{1} and \RNumb{2}, a priori information about the signal in the ``Radioastron'' mission can be obtained from webinet data or after preprocessing the Spektr-R satellite signal records using the JIVE algorithm (step ``swspectrometer'').

Using these sources, we know that the spectrum of the reference signal  $S^*(t) = A\cos\Phi^*(t)$ is concentrated (for  $\omega>0$) in a narrow band $(\omega^*_0 \pm \Delta\omega/2)$, where $\Delta\omega$ - frequency deviation due to the uneven motion of the SC in the orbit, $\omega^*_0 \simeq \omega_0$. Then, taking into account that $\varphi^*(t)$ - is a fairly close estimate of the informative parameter $\varphi(t,\mathbf{a})$, we have 
$$S(t,\mathbf{a}) \leftrightarrow S_\omega(\omega,\mathbf{a}) \neq0, \quad \mbox{when} \quad |\omega| \in (\omega^*_0 \pm \xi\Delta\omega/2),$$
$$ S_\omega(\omega,\mathbf{a}) - \mbox{signal spectrum}$$
and, therefore,
$S(t,\mathbf{a}) = A\cos\left[\omega_0t+\varphi(t,\mathbf{a})\right] = A\cos\left[\omega^*_0t+\varphi_D(t,\mathbf{a})\right],$
where 
$$\varphi_D(t,\mathbf{a}) = \varphi(t,\mathbf{a}) + (\omega_0 - \omega^*_0)t$$
$\xi$ - a scale factor of the order of 2-3, which is introduced to compensate for the uncertainty in the spectrum bandwidth.  \\
With this approach:
$$y(t) \rightarrow y_1(t) = \mbox{Re}\left[\tilde{y}_1(t)\exp\left\lbrace j\omega^*_0t\right\rbrace \right],  $$
where 
$$\tilde{y}_1(t) \leftrightarrow\tilde{y}_{1\omega}(\omega) = 2\tilde{y}_{\omega}(\omega^*_0+\omega), \quad |\omega|\le \xi\Delta\omega/2.$$
In further analysis, we will assume that the SNR in the $\Delta\omega_\xi$ band  turns out to be large:
$$q_1 = \frac{A^2}{2N\Delta f} \gg 1\quad (\Delta f = \frac{\Delta\omega}{2\pi}),$$
and, therefore, the phase of the signal is:
$$\theta_1(t) = arg\left\lbrace \tilde{y}_1(t)\right\rbrace  \approx \varphi_D(t,\mathbf{a}) + \varphi_n(t), $$
here $\varphi_n(t)$ - broadband gaussian noise with spectral density:
$$N_n(\omega) = \frac{2N}{A^2}, \quad |\omega|\le \xi\Delta\omega/2.$$ 
Using a priori information about the signal from the ``swspectrometer'', 
narrowing the bandwidth, we remove the trend in the frequency deviation zone.
The processing circuit for a phase detector can be represented as Fig 2

\begin{figure}[h!]
	\centering
	\includegraphics[width=1\linewidth]{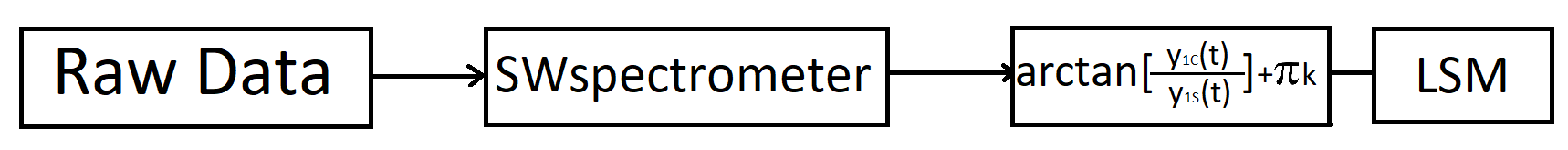}
	\caption{Data processing scheme in the "phase detector" mode: Raw Data - recording of the Spektr-R signal, SWspectrometer - spectrometer with a bandpass filter, Quadrature phase detection, LSM - regression analysis block.}
	\label{fig:scheme2}
	\centering
\end{figure}

The processing scheme for a phase detector can be represented as (\ref{scheme}).
Processing at the Fig. 2 is the extraction of quadrature components:
$$y_{1C}(t) = \mbox{Re}\ \tilde{y}_1(t), \quad y_{1S}(t) = \mbox{Im} \ \tilde{y}_1(t),$$
with the subsequent extraction of the signal's phase. Then the so-called ``Unwrap phase'' (taking into account its jumps $\pi k$) and the regression analysis are using the LSM (least squares method).
\begin{widetext}
\begin{eqnarray}\label{scheme}
\vec{y}(t)\rightarrow\left| \left|
\begin{array}{c}
y_{1C}(t)  \\
y_{1S}(t)   \\
\end{array}
\right|\right| \rightarrow\theta_1(t)=\arctan\left[ {\frac{y_{1C}(t)}{y_{1S}(t)}}\right]  + \pi k \rightarrow\overset{\overset{(\omega_0 - \omega^*_0)}{\downarrow}}{\mbox{LSM}}\rightarrow \mathbf{\hat{a}}.
\end{eqnarray}
\end{widetext}

Using the signal information from the swspectrometer, it is possible to remove the trend within the frequency deviation, practically narrowing the bandwidth to $\Delta\omega <5$ Hz.
Considering that for the signal model, linear in the estimated parameters:

$$\varphi_D(t,\mathbf{a}) = \sum_{k}a_kt^k + (\omega_0 - \omega^*_0)t,$$
the maximum likelihood estimates are effective \cite{linkk13} for SNR $\gg1$ we have:
$$<(\hat{a}_k - a_i)^2> \simeq \sigma^2_{ii},$$
where $\sigma^2_{ii}$ - diagonal elements of the inverse Fisher matrix. 

Simple calculations show that under the conditions of the ``Radioastron'' mission, SNR $>10^5$ are sufficient for the variance estimates to reach the level of the Rao-Cramer limit.
Thus  for sufficiently large SNR, the optimal discriminator can be replaced with a phase detector. The advantage of this method is a gain in processing time, as well as the fact that the estimates of informative parameters are not biased.

\section{Illustration of practical processing of selected sessions}
The primary processing of the received signals was carried out using JIVE algorithms \cite{linkk15} (codes) and data from standard frequency meters on the ``webinet'' website. Recording at Pushchino station was carried out in RDF (Radioastron Data Format).  Characteristics of the digital signal: the digitization frequency is 32 MHz, quantization by amplitude is 2 bits, a pure 8.4 GHz monochromatic signal is received by the video converter in the lower side band of about 6 MHz. At the first stage, the ``swspectrometer'' sub-program was used to plot the signal spectrum.
In this case, the information about  frequency deviation bandwidth is extracted from the spacecraft traffic data and the data from the webinet. On the basis of such (a priori) information, the sample of dividing the signal into stationary intervals is chosen so small that the frequency deviation does not affect the spectrum broadening. 
For a normal 1-way interval with a duration of 70 seconds we choose a split step under 1 second. The purpose of this stage is the most accurate determination of the frequency corresponding to the maximum of the spectral amplitude. After getting the spectra at each one-second interval, the dependence of the frequency on time is plotted, followed by the approximation of the trend by the least squares method.
\begin{figure}[h!]
	\centering
	\includegraphics[width=1\linewidth]{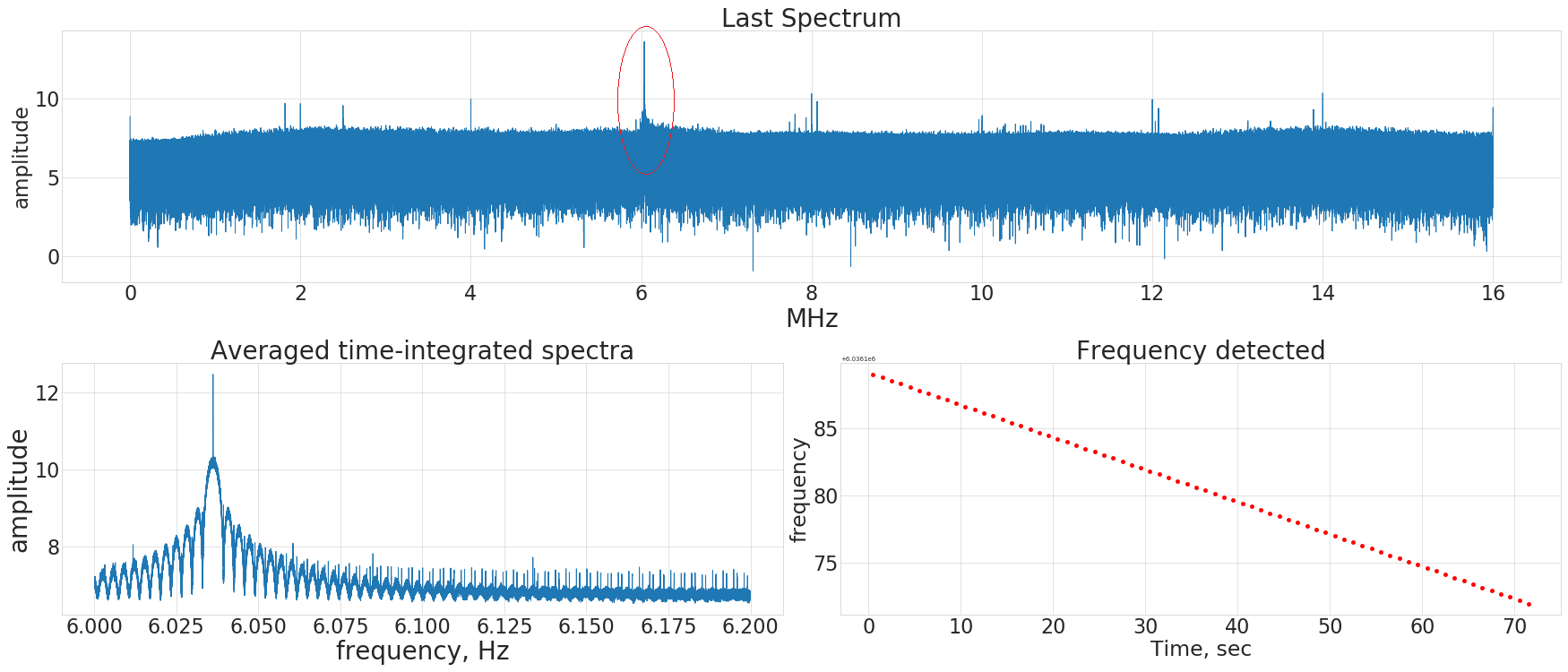}
	\caption{The upper graph shows one of the spectra plotted on a one-second interval, at a frequency of about 6 MHz; the main tone is highlighted by which the signal frequency is determined. The lower-left graph shows the averaged signal spectrum over all intervals. The lower right graph shows the dependence of frequency on time, each point corresponds to a frequency determined by the maximum in the spectrum.}
	\label{fig:CalcCPP}
	\centering
\end{figure}
After the ``swspectrometer'' stage, we have the phase and frequency polynomials for the signal research.
Using the phase polynomial as a first approximation, we remove the main trend in frequency deviation using the ``SCtracker'' program. At the output, we receive a signal in a very narrow band $<5$ Hz, which we then process using the phase detection method.

\begin{figure}[h!]
	\centering
	\includegraphics[width=1\linewidth]{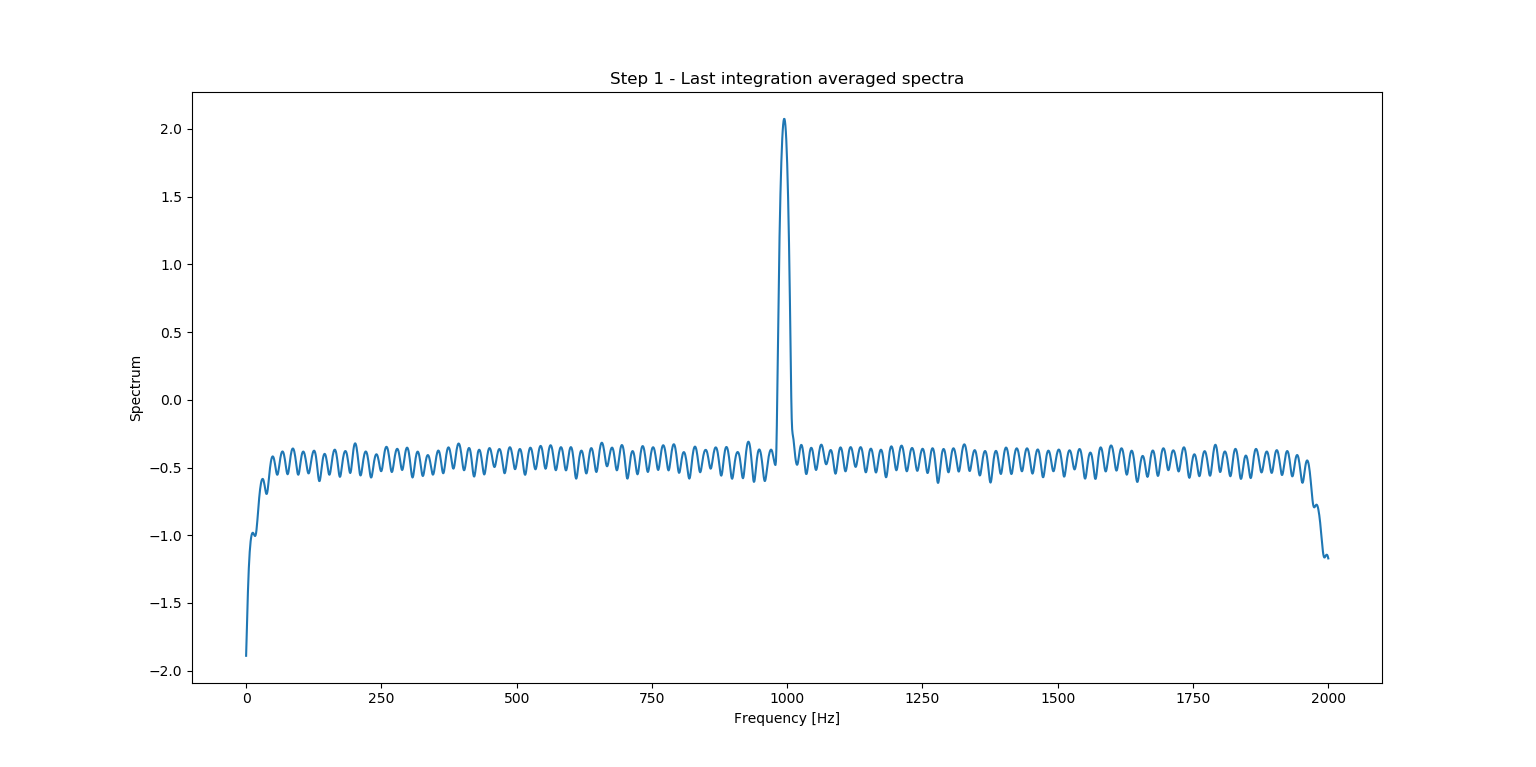}
	\caption{The spectrum of the signal after processing by SCtracker. The spectrum is concentrated in a narrow band, the signal can be considered stationary}
	\label{fig:SCtrack}
	\centering
\end{figure}
For a stationary signal, one can use the standard Hilbert transform programs with the subsequent separation of the signal phase. 
\begin{figure}
	\centering
	\includegraphics[width=1\linewidth]{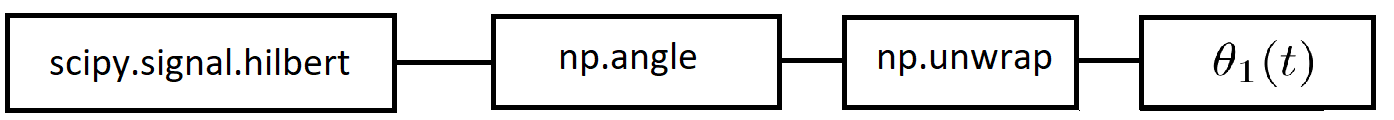}
	\caption{Plotting a phase using the numpy (np) and scipy program library}
	\label{fig:pic3}
	\centering
\end{figure}
Further, the evolution of the signal phase is approximated by the least squares method. The regression line equation gives the values of the coefficients of the phase polynomial.

To calculate the error of the coefficients of the phase polynomial, the Rao-Cramer inequality is used, which requires the calculation of the Fisher information matrix. When calculating the elements of the inverse Fisher matrix, SNR data obtained from the ASC (Astro Space Center) for a particular gravity session, as well as an independent calculation of the SNR from real GTS records of Pushchino, are taken into account. In order of magnitude, both SNR coincide, which makes it possible to fairly reliably calculate the variance of the coefficients of the phase polynomial:

\begin{table}[h!]
\caption{ Accumulation of all gravitational sessions would allow to reach the level of epsilon estimate $\epsilon \sim 3\cdot10^{-5}$}
\begin{tabular}{|c|c|c|c|c|c|}
\hline
session code & SNR 1w	& SNR 2w & raw redshift, Hz & Rao-Cramer error, Hz &  relative accuracy\\
\hline
raks17bb & 271985 & 795103 & 4.64101085 & 0.00041514 & 4.9421e-14 \\
\hline
raks17bh &330866 &540078 &5.4447583 &0.0037639 & 4.4808e-14  \\
\hline
raks17bj & 355954&508597 &5.00294562 &0
00036288 &4.3201e-14  \\
\hline
raks17bc & 218414& 443638&4.40564635 &0.00046326 & 5.55150e-14 \\
\hline
raks17bk &346025 &550515 &5.60846747 &0.00036805 &4.3816e-14  \\
\hline
\end{tabular}	
\end{table}

\section{Conclusion}
In this paper, a technique was presented that can be used to measure the gravitational frequency shift of radio communication signals with an accuracy exceeding the capabilities of conventional frequency meters (webinet data). At the same time, estimates of the carrier frequency of communication signals using the specialized JIVE algorithm are also limited in accuracy due to short signal recording intervals (1-3 min) in a fixed synchronization mode with reference frequency standards (1w, 2w). In our proposed processing procedure, the measurement of the gravitational shift (as the frequency difference in modes (1w, 2w)) occurs as a result of phase detection after bandpass frequency filtering of the communication signal, using its extremely narrow spectral band. The effective bandwidth (as a priori information) is determined using the first stage of the JIVE algorithm (swspectrometer stage). The estimation of the measurement accuracy of the gravitational shift is performed by calculating the Rao-Cramer limit, which is based on the large signal-to-noise ratios in the sessions of the ground station with the satellite (SNR = $10^5-10^6$). \\
Testing the proposed technique on the data of a specific communication session between the Pushchino station and the Spektr-R apparatus showed its efficiency. It is expected that, with its help, the processing of records of gravitational sessions accumulated during the Radioastron mission will show the correspondence between the GR theory and the experiment with an accuracy of $10^{-4} - 10^{-5}$, i.e. at the level of the results obtained in measurements with Galileo satellites \cite{linkk4},\cite{linkk5}. \\ 

\section{Acknowledgments}
The authors express their gratitude to the members of the international group of researchers on the ``Radioastron'' project; N. Bartel, N. Nunes (York Univ.), L. Gurvitz (JIVE), D. Litvinov (ASC), Zakhvatkin (AMI) for helpful discussions of this technique. \\ 
This work was supported by the grant RFBR 19-29-11010.

\end{document}